\documentclass{aa_edited}  

\usepackage{graphicx,txfonts,booktabs,natbib,mathtools,microtype,amssymb,amsmath,xspace,siunitx}
\usepackage[usenames,dvipsnames]{color} 
\usepackage[normalem]{ulem}
\usepackage[caption=false]{subfig} 
\usepackage{hyperref}
\usepackage{orcidlink}

\def\apjl{{Astrophys.~J.~Lett.}}
\def\apjs{{Astrophys.~J.~Supp.}}
\def\ao{{Appl.~Opt.}}

\def\aap{{Astron.~Astrophys.}}

\newcommand{\lsim}{\mathrel{\rlap{\lower 3pt \hbox{$\sim$}} \raise 2.0pt \hbox{$<$}}}
\newcommand{\gsim}{\mathrel{\rlap{\lower 3pt \hbox{$\sim$}} \raise 2.0pt \hbox{$>$}}}
\definecolor{linkcolor}{rgb}{0.0,0.3,0.5}
 \hypersetup{
     colorlinks=true,
     linkcolor=linkcolor,
     filecolor=linkcolor,
     citecolor = linkcolor,      
     urlcolor=linkcolor,
     }

\newcommand{\ssim}{\mathchar"5218\relax\,}

\def\surname{{NRSur7dq4Remnant}\xspace}
\def\quasar{{3C~186}\xspace}
\def\msun{M_{\odot}}

\newcommand{\bham}{{School of Physics and Astronomy \& Institute for Gravitational Wave Astronomy, University of Birmingham, Birmingham, B15 2TT, United Kingdom}}
\newcommand{\milan}{{Dipartimento di Fisica ``G. Occhialini'', Universit\'a degli Studi di Milano-Bicocca, Piazza della Scienza 3, 20126 Milano, Italy}}
\newcommand{\infn}{{INFN, Sezione di Milano-Bicocca, Piazza della Scienza 3, 20126 Milano, Italy}}
\newcommand{\inaf}{{INAF Osservatorio Astronomico di Brera, Via Bianchi 46, 23807 Merate, Italy}}

\begin{document} 

\title{Astrophysical and relativistic  modeling of the\\recoiling black hole candidate in quasar \quasar}

\titlerunning{Recoiling black hole in quasar \quasar}

\author{
Matteo Boschini$\,$\orcidlink{0009-0002-5682-1871}\inst{1,2} 
\and Davide Gerosa$\,$\orcidlink{0000-0002-0933-3579}\inst{1,2,3}        
\and Om Sharan Salafia$\,$\orcidlink{0000-0003-4924-7322}\inst{2,4}
\and Massimo Dotti$\,$\orcidlink{0000-0002-1683-5198}\inst{1,2}
}

\authorrunning{Boschini et al.}

\institute{\milan \and \infn \and \bham \and \inaf
\\[3pt] E-mail: \href{mailto:m.boschini1@campus.unimib.it}{m.boschini1@campus.unimib.it}    
}

\abstract{The compact object in quasar 3C 186 is one of the most promising recoiling black hole candidates, exhibiting both an astrometric displacement between the quasar and the host galaxy as well as a spectroscopic shift between broad and narrow lines. 3C 186 also presents a radio jet that, when projected onto the plane of the sky, appears to be perpendicular to the quasar-galaxy displacement. Assuming a gravitational-wave kick is indeed responsible for the properties of 3C 186 and using state-of-the-art relativistic modeling, we show that current observations allow for exquisite modeling of the recoiling black hole. Most notably, we find that the kick velocity and the black hole spin are almost collinear with the line of sight and the two former vectors appear perpendicular to each other only because of a strong projection effect. The targeted configuration requires substantial fine-tuning: while there is a region in the black hole binary parameter space that is compatible with 3C 186, the observed system appears to be a rare occurrence. Using archival radio observations, we explored different strategies that could potentially confirm or rule out our interpretation. In particular, we developed two observational tests that rely on the brightness ratio between the approaching and receding jet as well as the asymmetry of the jet lobes. While the available radio data provide loose constraints, deeper observations have the unique potential of unveiling the~nature~of~3C~186. %
}
\keywords{black-hole physics -- quasars: supermassive black holes -- relativistic processes}

\maketitle

\section{Introduction}

The formation of supermassive black holes (BHs) likely involves initial seeds undergoing accretion and merger episodes \citep{2021NatRP...3..732V}. Empirical relations between the BH mass, the stellar dispersion velocity, and the mass of the galactic bulge \citep{2013ApJ...764..184M} suggest the occurrence of common evolutionary processes by which the growth of the BH is linked to that of its host galaxy and influences its properties. %
This interplay is most evident for active galactic nuclei (AGNs): material falling toward the bottom of the galaxy potential well forms an accretion disk around the supermassive BH that, in turn, shapes the surrounding environment.

Most galaxies are expected to undergo at least one major merger during their evolution \citep{2021MNRAS.501.3215O}, and this is especially true for the brightest cluster galaxies sitting at the center of clusters \citep{2015MNRAS.446...38G}.
Galactic mergers plays a fundamental role in the formation of cosmic structures %
and provide new influxes of gas, ultimately leading to an increased rate of star formation and the powering of AGNs. Following a galaxy merger, the supermassive BHs hosted by the two galaxies are expected to form binaries and eventually merge  --- a process that likely involves a variety of mechanisms, including dynamical friction, stellar scattering, gas-assisted migration, and gravitational wave (GW) emission \citep{1980Natur.287..307B, 2014SSRv..183..189C}.
A potential strategy for inferring the occurrence of supermassive BH mergers besides direct GW probes \citep{2017arXiv170200786A} is to detect post-merger electromagnetic signatures \citep[e.g.,][]{2004ApJ...606L..17M, 2007ApJ...666L..13B,2016MNRAS.456..961B}.%

Anisotropic emission of GWs during the late BH inspiral and merger imparts a proper velocity, or ``kick,'' to the BH remnant~\citep{1961RSPSA.265..109B, 1962PhRv..128.2471P, 1973ApJ...183..657B}. The kick velocity is independent of the total mass of the binary and is typically  $\mathcal{O}(100\,\si{km/s})$ (see, e.g., \citealt{2018PhRvD..97j4049G}). %
However, some fine-tuned spin configurations give rise to ``superkicks'' of  $\mathcal{O}(1000\,\si{km/s})$
\citep{2007PhRvL..98w1101G,2007PhRvL..98w1102C}. Such velocities are comparable to, if not higher than, the escape speed of the galactic host itself \citep{2004ApJ...607L...9M}.%

There are two main observational strategies used to identify recoiling supermassive BHs; both assume that some of the nearby material %
 remains bound to the BH after the merger  \citep{2007PhRvL..99d1103L, 2008ApJ...687L..57V}: %
\begin{itemize}
\item Astrometric surveys are used to identify off-nuclear AGNs, targeting objects with a projected position that is somewhat displaced from the galactic center \citep[e.g.,][]{2004ApJ...606L..17M}.%
\item Spectroscopic searches for recoiling BHs target the predicted Doppler shift between the so-called broad lines, which originate from material bound to the BH, and the narrow lines, which instead track the host galaxy \citep[e.g.,][]{2007ApJ...666L..13B}.%
\end{itemize}
Even though these signatures are not unique, several recoiling supermassive BH candidates have been proposed  %
\citep{2008ApJ...678L..81K, 2010ApJ...717L...6B, 2010ApJ...717L.122R, 2012ApJ...752...49C, 2012ApJ...759...24S, 2014ApJ...789..112C, 2014MNRAS.445..515K, 2014ApJ...795..146L, 2015A&A...580A..11M,
2016ApJ...829...37B, 2016ApJS..222...25S, 2017A&A...600A..57C, 2017ApJ...840...71K, 2017ApJ...851L..15K, 2018ApJ...861...51K,
2018ApJ...863..149P, 2018MNRAS.475.5179S, 2021ApJ...909..141P, 2021MNRAS.503.1688H}%
, some of which have since been ruled out \citep{2014MNRAS.445.1558D, 2024ApJ...961...19L}.
The most promising candidate to date is arguably the quasar \quasar~\citep{2017A&A...600A..57C, 2018ApJ...861...56C}. This source  presents both an evident astrometric displacement of $\ssim 10$ kpc between the quasar and the isophotal center of the galaxy stellar distribution and a $\ssim 2000$ km/s Doppler shift between broad and narrow  lines. %
 Using submillimeter interferometry, \cite{2022A&A...661L...2C} reported the detection of a large reservoir of molecular gas ($\gtrsim 10^{10} \msun$) orbiting the center of the observed galaxy; at the same time, cold gas was not detected at the location and redshift of the quasar, in line with the recoiling BH scenario.\footnote{The same test applied to the %
  candidate first identified by \cite{2008ApJ...678L..81K} resulted in the identification of large amounts of cold molecular gas co-located with the quasar, thus ruling out the recoiling hypothesis \citep{2014MNRAS.445.1558D}.}

\cite{2022ApJ...931..165M} point out that the projected direction of the jet (which presumably tracks the spin of the BH) in \quasar is essentially perpendicular to the projected separation between the AGN and the photometric center of the galaxy (which presumably tracks the direction of the kick). If \quasar is a post-merger BH, its kick and spin appear perpendicular to each other. This is puzzling, at least at first sight, because general relativity strongly predicts that large kicks of $\mathcal{O}(1000)~\si{km/s}$ tend to be parallel to the BH spin \citep{2018PhRvD..97j4049G}.  
In this paper we show that this new piece of observational evidence together with detailed relativistic modeling allows for an unprecedented characterization of \quasar. Compared to the previous study by \cite{2017ApJ...841L..28L}, our analysis includes a complete modeling of all vector components of spin and kick, which are crucial to solving the geometry of the system. Our statistical methods are also more advanced, similar to those used by \cite{2023ApJ...945L..18P} for the recoiling BH  candidate E1821+643 \citep{2016ApJS..222...25S}. The key advantage of studying \quasar is the additional information provided by the projected angle between the spin and the kick.

In particular, we fully resolve the intrinsic orientation of the system. For \quasar to be compatible with a recoiling BH, we find that the spin, the kick, and the line of sight must be close to collinear. The BH spin and kick thus appear perpendicular to each other only because of a prominent projection effect, while in reality they are actually almost parallel. Assuming broad uninformative priors, we show that, if \quasar is indeed a recoiling BH, it must be a very unusual one (though its identification might be favored by selection biases). In light of our findings, we formulated additional observational tests that could unveil the true nature of \quasar, ultimately confirming or ruling out the recoiling BH hypothesis. In particular, the spatial and brightness asymmetry of the approaching and receding radio jet can be used to measure the viewing angle of the jet, which is presumably being launched along the direction of the BH spin.

 Our paper is organized as follows. In Sect. \ref{sec:two} we review the current observations of  \quasar. In Sect. \ref{sec:three} we present our statistical modeling and the resulting constraints on the properties of \quasar. In Sect. \ref{sec:four} we discuss the astrophysical implications of our results and present potential observational tests using radio data. 
 In Sect.~\ref{sec:five} we present some concluding remarks. Some additional sanity checks are provided in Appendix~\ref{secapp}.

\section{Quasar \quasar}
\label{sec:two}
The radio-loud quasar  
\quasar is located in a well-studied cluster of galaxies at redshift $z=1.06$ %
\citep{2010MNRAS.405.2302H}. %
 \quasar is powered by a BH of mass $\ssim 3-6\times 10^9\,\si{M_{\odot}}$ inferred using both spectroscopic and photometric data %
\citep{2017A&A...600A..57C}.
Images from the \textit{Hubble} Space Telescope 
 with the Wide Field Camera 3
  and the Advanced Camera for Surveys 
  \citep{2017A&A...600A..57C, 2022ApJ...931..165M} show that the AGN is located at a projected separation %
of $11.1 \pm 0.1\,\si{kpc}$ away from the photometric center of the host galaxy, coinciding with the kinematic center of its molecular gas \citep{2022A&A...661L...2C}. Spectroscopic data were obtained with the \textit{Hubble} Space Telescope Faint Object Spectrograph \citep{2017A&A...600A..57C}, the Sloan Digital Sky Survey \citep{2017A&A...600A..57C}, the Keck/OSIRIS Integral Field Units \citep{2018ApJ...861...56C} and the Northern Extended Millimeter Array \citep{2022A&A...661L...2C}.  In particular, \cite{2017A&A...600A..57C} reported a velocity shift between the broad and narrow emission lines of $-2140 \pm 390\,\si{km/s}$, where the minus sign indicates that the quasar is receding slower %
than the galaxy.

The presence of both astrometric and spectroscopic offsets drew attention on \quasar as a prominent recoiling BH candidate. %
As discussed at length by \cite{2017A&A...600A..57C}, there are other plausible scenarios. AGN disk features such as extreme emitters or winds can explain some of the shifted lines.  Regarding the apparent offset, the observed system could be a superposition of two galaxies at different redshifts, where the one hosting the AGN is very compact. While these properties could explain the observations of \quasar, they also require ad hoc assumptions and/or the presence of additional undetected structure (see \citealt{2017A&A...600A..57C} for a thorough discussion).

The radio jet of \quasar was observed by the \textit{Karl Jansky} Very Large Array (VLA) at 8.4 GHz. Data are publicly available at the National Radio Astronomy Observatory archive with project ID~AA129 (PI: C.~E.\ Akujor). Earlier, higher-resolution observations from the European Very Long Baseline Interferometry Network (EVN) and the Multi-Element Radio Linked Interferometer Network (MERLIN) at 1.7 GHz have been reported by \cite{1991MNRAS.250..225S}. Unfortunately, the original data are not available anymore (Spencer 2024, private communication); subsequent MERLIN observations at a higher frequency, with a comparable resolution, are present in the MERLIN archive and were kindly shared with us (Williams 2024, private communication). Since these observations are less sensitive, and hence produce shallower constraints on the quantities of interest, in the following we make use of the VLA data (which we reduced; see Sect.~\ref{jetorientation}) and some summary information from the EVN+MERLIN observations as reported in \cite{1991MNRAS.250..225S}.

\section{Relativistic modeling}
\label{sec:three}

\begin{figure}
        \centering
        \subfloat{\includegraphics[trim={20.8cm 11cm 21cm 8cm}, clip, width=0.5\textwidth]{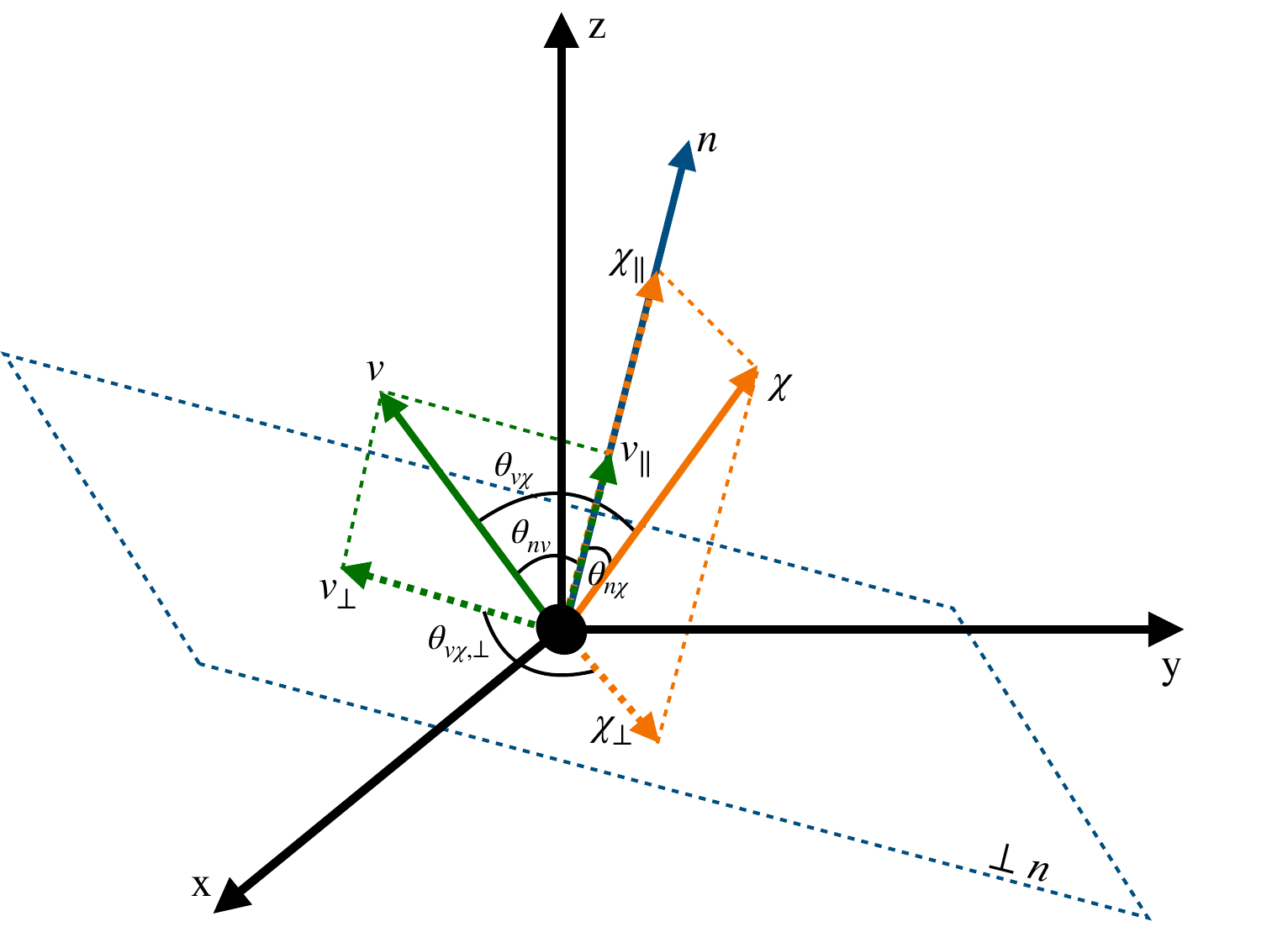}}
        \caption{Schematic representation of the targeted geometry. The black dot in the center indicates the post-merger BH remnant. The spin ($\chi$) is indicated in  orange, the kick velocity ($v$) is indicated in green, and the line of sight ($n$) is indicated in blue. Subscripts $\parallel$ and $\perp$ indicate projections with respect to the line of sight. %
        }
        \label{fig:arrows}
        \end{figure}

We test whether the observed properties of \quasar can be explained by a GW kick as predicted by general relativity. In other words: is there a BH binary that produces a post-merger BH compatible with the observations? A schematic representation of the geometry of the system is provided in Fig.~\ref{fig:arrows}.

\subsection{Bayesian sampling}

The properties of the putative pre-merger binary and the direction of the line of sight need to satisfy the  following two constraints:
First, the kick velocity projected along the line of sight is
\begin{equation}
v_{\parallel} = -2140 \pm 390\,\si{km/s}
,\end{equation}
as inferred from spectroscopic data \citep{2017A&A...600A..57C}.  Second, te angle between the kick and the spin projected onto the plane of the sky is  
\begin{equation}
\theta_{v\chi,\perp} = \arccos\bigl(\hat{\boldsymbol{\chi}}_{\perp} \cdot \hat{\boldsymbol{v}}_{\perp}\bigr) =  \ang{90} \pm \ang{10}
,\end{equation}
as first pointed out by \cite{2022ApJ...931..165M}. %

 While 
\cite{2022ApJ...931..165M} do not explicitly provide an error budget for $\theta_{v\chi,\perp}$, they report significant digits up to $\ang{1}$. We thus opted for a conservative uncertainty of $\ang{10}$. In Appendix \ref{secapp} we show that assuming an error on $\theta_{v\chi,\perp}$ of $\ang{1}$ instead of $\ang{10}$ returns  probability distributions that are largely indistinguishable.

We present a Bayesian analysis using a Monte Carlo Markov chain (MCMC) sampler to identify the regions of the parameter space that are compatible with the observations.  We assumed a binary progenitor on a quasi-circular orbit characterized by a mass ratio $q={m_2}/{m_1}$, where $m_1$ ($m_2$) is the mass of the heavier (lighter) BH, and dimensionless spin vectors $\boldsymbol{\chi}_{1,2}$. They are described by magnitudes $\chi_{1,2}\in[0,1)$, polar angles $\theta_{1,2}$ between the spins and the orbital angular momentum, and azimuthal angles $\phi_{1,2}$ measured in the orbital plane. In the following, we often report the difference $\Delta\phi=\phi_1-\phi_2$ because it tracks the occurrence of large recoils \citep{2008PhRvD..77l4047B}. We predicted the spin and kick of the post-merger BH using the machine-learning model \surname \citep{2019PhRvR...1c3015V,2019PhRvL.122a1101V}, %
which is trained on numerical-relativity simulations of merging BHs. We adopted their conventions and defined the spin angles at a time $t=-100 G (m_1+m_2)/c^3$ before BH merger. %

We sampled the parameter space using \textsc{emcee}  \citep{2013PASP..125..306F}.  We used a bivariate Gaussian likelihood,
\begin{equation}
\mathcal{L}(d | \boldsymbol{x}) = \mathcal{N}( \boldsymbol{x} | \boldsymbol{\mu},\boldsymbol{\Sigma})
,\end{equation}
where $d$ stands for the data and
\begin{equation}
\boldsymbol{x} =\begin{pmatrix} v_\parallel \\  \theta_{v\chi,\perp} \end{pmatrix}\,.
\end{equation}
The likelihood mean and covariance are 
\begin{equation}
\label{musigma}
\boldsymbol{\mu} =\begin{pmatrix} -2140\; {\rm km/s} \\ \ang{90}\end{pmatrix}\,,
\qquad \quad
\boldsymbol{\Sigma} =\begin{pmatrix} 390\; {\rm km/s} & 0  \\ 0 & \ang{10}\end{pmatrix}^2\,.
\end{equation}

We set uninformative priors on all our target parameters. %
Specifically, we sampled the mass ratio $q\in [1/6,1]$ uniformly (the lower bound was set by the validity of the \surname model and we have verified this does not affect our results), the dimensionless spin magnitudes $\chi_{1,2}\in [0,1)$ uniformly, and the spin directions isotropically. The direction of the line of sight is parametrized by two angles $\theta_{n}$ and $\phi_{n}$, %
 and is also assumed to be isotropic. The fact that these angles are measured in the same reference frame of the spins (i.e., $\theta_{n}=\ang{0}$) implies that the line of sight is parallel to the pre-merger orbital angular momentum. %

The MCMC sampler converges to a region of the parameters space of the progenitor binaries that is compatible with \quasar being a recoiling remnant. That said, our MCMC run shows a low acceptance rate of $9.6\%$ %
and a considerable autocorrelation length $\ssim2\times10^4$ (where we quote the maximum value across 9 parameters) out of $5\times10^5$ steps for each walker.  %
This indicates an intrinsic difficulty at locating the maximum of the posterior under our uninformative prior.

\begin{figure*}[]
        \centering
        \subfloat{\includegraphics[width=\textwidth]{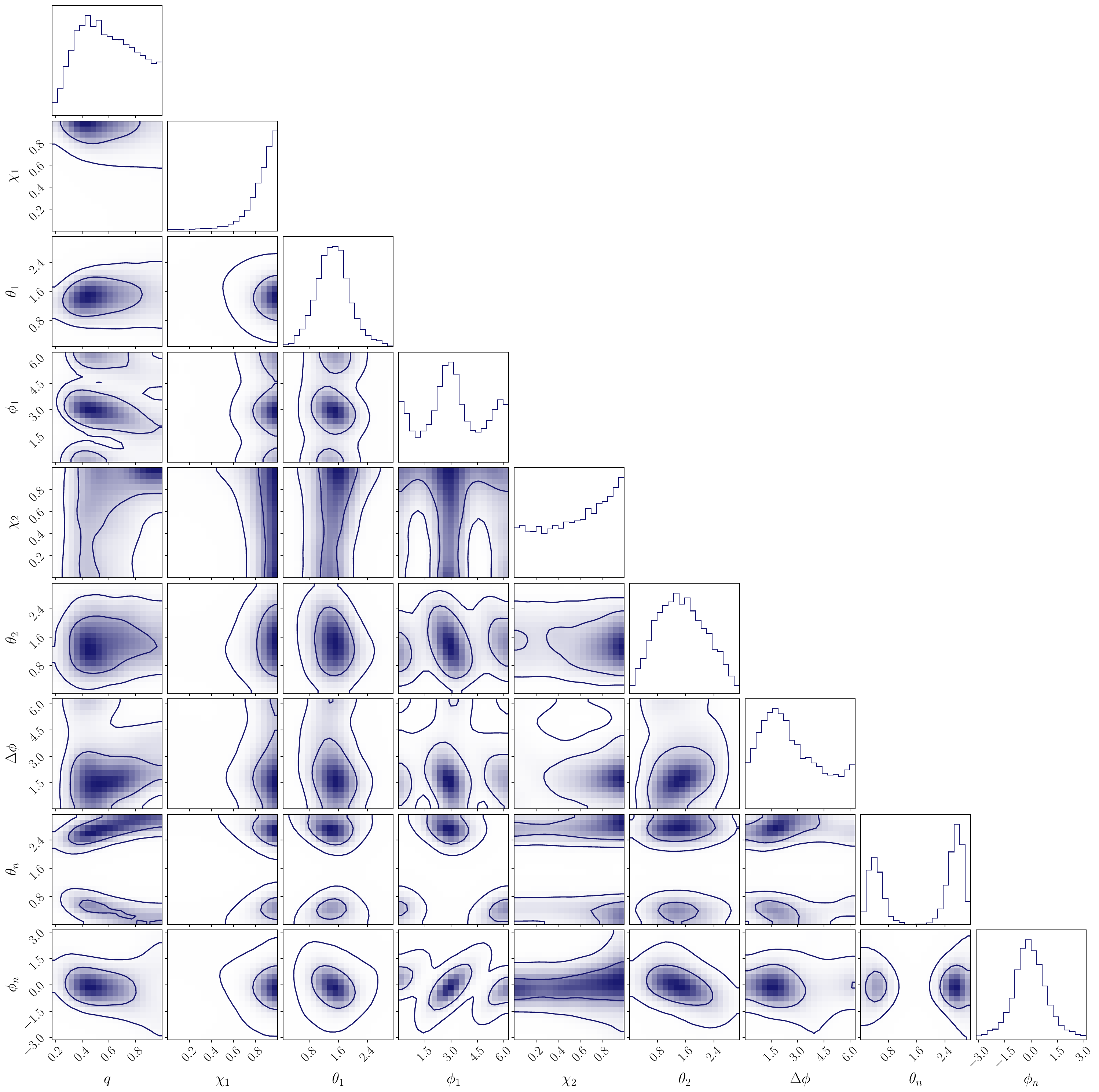}}
        \caption{Joint posterior distributions of the parameters describing the binary progenitors of \quasar. We show one- and two-dimensional marginal posteriors for the mass ratio ($q$), spin magnitudes ($\chi_{1,2}$), spin inclination angles ($\theta_{1,2}$), the in-plane azimuthal angle of the heavier BH spin ($\phi_1$), the difference between the azimuthal angles ($\Delta\phi$), polar observation angles ($\theta_{n}$), and the azimuthal observation angle ($\phi_{n}$). Contours indicate  $90\%$ and $50\%$ credible regions.} 
        \label{fig:cornerplot}
\end{figure*}

Figure \ref{fig:cornerplot} shows the posterior distribution of all the sampled parameters. Remarkably, this is bimodal, with two configurations that are compatible with the observed properties. The distinguishing factor is $\theta_{n}$, the angle between the line of sight and the $z$-axis that traces the direction of the orbital angular momentum before merger. %
Medians and $90\%$ credible intervals of some parameters are listed in Table~\ref{tab:MCMCest}, where we separate the two modes of the distribution. 

We also performed an additional run where we omitted the constraint on $\theta_{v\chi,\perp}$ and only considered that on $v_\parallel$; the results are reported in Appendix \ref{secapp}. 
The posterior is visibly different, and in particular prefers binaries with larger mass ratios; the direction of the line of sight is also affected. This
suggests that the jet direction is indeed important to constrain the properties of the system.%

\begin{table}\centering
\setlength{\tabcolsep}{6pt}
\renewcommand{\arraystretch}{1.5}
\begin{tabular}{l@{\extracolsep{0.2cm}}l|cc}
  && $\ang{0} \!<\! \theta_{n} \!<\! \ang{90}$ & $\ang{90} \!<\! \theta_{n} \!<\! \ang{180}$ \\
  \hline \hline
  $q$    &                     & $0.58^{\,+0.37}_{\,-0.29}$ & $0.57^{\,+0.37}_{\,-0.29}$ \\
  $\chi_1$       &             & $0.89^{\,+0.10}_{\,-0.43}$ & $0.89^{\,+0.10}_{\,-0.39}$ \\
  $\theta_1$ & [deg]       & $83.3^{\,+50.4}_{\,-42.7}$ & $82.8^{\,+43.7}_{\,-42.3}$ \\
  $\phi_1$ & [deg]       & $344.8^{\,+82.1}_{\,-83.0}$ & $165.9^{\,+87.1}_{\,-82.5}$ \\
  $\chi_2$ &                    & $0.63^{\,+0.35}_{\,-0.56}$ & $0.59^{\,+0.38}_{\,-0.53}$ \\
  $\theta_2$ & [deg]            & $83.0^{\,+69.2}_{\,-57.5}$ & $83.4^{\,+67.8}_{\,-58.3}$ \\
  $\Delta\phi$ & [deg]      & $131.7^{\,+204.2}_{\,-108.0}$ & $132.6^{\,+201.8}_{\,-111.3}$\\ 
  $\theta_n$ & [deg]            & $24.3^{\,+23.3}_{\,-15.3}$ & $155.5^{\,+15.8}_{\,-23.6}$ \\
  $\phi_n$ & [deg]            & $-5.3^{\,+102.5}_{\,-92.8}$ & $-5.4^{\,+105.0}_{\,-96.4}$ \\
  \hline 
  $v_{\parallel}$ & [km/s]        & $ -1740^{\,+595}_{\,-657}$ & $-1740^{\,+574}_{\,-640}$ \\
  $\theta_{v\chi,\perp}$ & [deg]   & $87.2^{\,+17.5}_{\,-16.7}$ & $93.0^{\,+16.4}_{\,-15.8}$ \\
    \hline 
    $v$ & [km/s] & $1759^{\,+595}_{\,-657}$ & $1759^{\,+571}_{\,-635}$ \\
    $v_{\perp}$ & [km/s] & $214^{\,+236}_{\,-154}$ & $213^{\,+224}_{\,-157}$ \\
    $\theta_{v\chi}$ & [deg] & $10.8^{\,+9.5}_{\,-5.7}$ & $169.4^{\,+5.6}_{\,-10.4}$ \\
    $\theta_{n\chi}$ & [deg] & $7.3^{\,+9.5}_{\,-5.3}$ & $172.8^{\,+5.1}_{\,-9.9}$ \\
    $\theta_{nv}$ & [deg] & $7.1^{\,+8.7}_{\,-5.1}$ & $7.1^{\,+8.6}_{\,-5.1}$
\end{tabular}
\medskip
\caption{Medians and $90\%$ credible intervals of the marginalized distributions of some of our parameters. We present results for the two modes of the joint posterior, which we define using the inclination angle $\theta_{n}$. We first report the parameters that directly enter our sampling algorithm: mass ratio $q$, spin magnitudes $\chi_{1,2}$, spin tilts $\theta_{1,2}$, spin azimuthal angles $\phi_1$ and $\Delta\phi=\phi_2-\phi_1$, and direction of the line of sight $\theta_n$ and $\phi_n$. %
We also build posterior distributions for the quantities that directly enter our likelihood, namely the projected kick velocity  $v_{\parallel}$ and the projected spin-kick angle $\theta_{v\chi,\perp}$. Finally, we show the measured values of some other derived parameters: kick magnitude $v$, projected recoil velocity $v_{\perp}$, spin-kick angle $\theta_{v\chi}$, observer-spin angle $\theta_{n\chi}$, and observer-kick angle $\theta_{nv}$.}%
\label{tab:MCMCest}
\end{table}

\subsection{Progenitors of \quasar}
BH remnants recoil because of anisotropic emission of GWs, which in turn is due to asymmetries between the masses and spins of the two merging BHs \citep{2018PhRvD..97j4049G}. %
Our reconstruction shows evident constraints in the parameter space of the putative BH binary that generated \quasar. Achieving recoils of $O(1000)\,\si{km/s}$ requires  %
 configurations with mass ratios close to unity ($q\sim 1$),  spin magnitudes close to their maximum values ($\chi_{1,2}\sim 1$), and spin directions that are sufficiently misaligned with respect to the orbital angular momentum ($\theta_{1,2}\gtrsim  \ang{50}$) and antiparallel to each other ($\Delta\phi\sim \ang{180}$;~\citealt{2007PhRvL..98w1101G,2007PhRvL..98w1102C,2011PhRvL.107w1102L}). %
Moreover, when the recoil velocity is large, its direction is largely parallel to the pre-merger orbital angular momentum \citep{2018PhRvD..97j4049G},  which in turn tracks the direction of the remnant spin. That is, we expect $\theta_{v\chi}\sim \ang{0}$ or \ang{180}.

The results of our MCMC run imply a kick of $\ssim 1700$ km/s, which is large but relatively far from the maximum kick achievable ($\sim 5000$ km/s; \citealt{2011PhRvL.107w1102L}). For this reason, the sampler favors binaries with somewhat asymmetric masses ($q\sim 0.6$) and with a high-spinning primary BH ($\chi_1\sim 0.9$). Constraints on the secondary spin magnitude are naturally weaker for asymmetric binaries because the ratio between the spin angular momenta is $q^2\chi_2/\chi_1$.
The spins $\boldsymbol{\chi}_1$ and $\boldsymbol{\chi}_2$ do not lie onto the orbital plane ($\theta_1\sim\theta_2\sim\ang{80}$), which indicates an interplay between the so-called superkicks (which occur for $\theta_i\sim \ang{90}$; \citealt{2007PhRvL..98w1101G,2007PhRvL..98w1102C}) and hang-up kicks (which occur for $\theta_i\sim \ang{50}$; \citealt{2011PhRvL.107w1102L}). 

We find a strong correlation between the modes of $\theta_n$ and those of $\phi_1$ (see  Fig. \ref{fig:cornerplot}). This behavior is set by the strong dependence of BH kicks on the orbital phase at merger. A detailed investigation on this point was first presented by \cite{2008PhRvD..77l4047B} and repeated with updated models by \cite{2018PhRvD..97j4049G}. In brief, for systems that can excite large kicks, the center of mass of the binary oscillates during the inspiral because of the frame dragging of the two BHs. The merger halts this oscillation, which implies the resulting kick follows a roughly sinusoidal behavior with the orbital phase at merger, which is in turn degenerate with the azimuthal spin projections. In this case, the two $\phi_1$ modes correspond to recoils that are qualitatively similar but in opposite directions (the ``top'' and the ``bottom'' of the center-of-mass oscillation), so this angle is highly correlated with $\theta_n$.

\subsection{Geometry of \quasar}

The spin and the kick of \quasar are either almost parallel ($\theta_{v\chi}\sim \ang{0}$ for the mode with $\ang{0} \!<\! \theta_{n} \!<\! \ang{90}$) or almost antiparallel ($\theta_{v\chi}\sim \ang{180}$ for the mode with $\ang{90} \!<\! \theta_{n} \!<\! \ang{180}$) %
to each other (see Table~\ref{tab:MCMCest}).
This does not contradict the observation that the angle $\theta_{v\chi,\perp}$ between the direction of the astrometric offset and that of the radio jet needs to be $\ssim \ang{90}$, which is indeed is recovered correctly in our sampling run. We are here breaking some of the intrinsic degeneracies of the system. Our key finding is that object \quasar is oriented such that the observer is nearly (anti)aligned with both the remnant spin and kick vectors. The kick ($\boldsymbol{v}$), the spin  ($\boldsymbol{\chi}$), and the line of sight ($\boldsymbol{n}$) are almost collinear; the projections of the kick and spin vectors into the plane of the sky are some small vectorial components, oriented in a way that they appear perpendicular to each other. As a consequence, we find $v \sim |v_{\parallel}| \gg v_{\perp}$ (see Table~\ref{tab:MCMCest}).%

\begin{figure}[]
        \centering
        \subfloat{\includegraphics[width=0.5\textwidth]{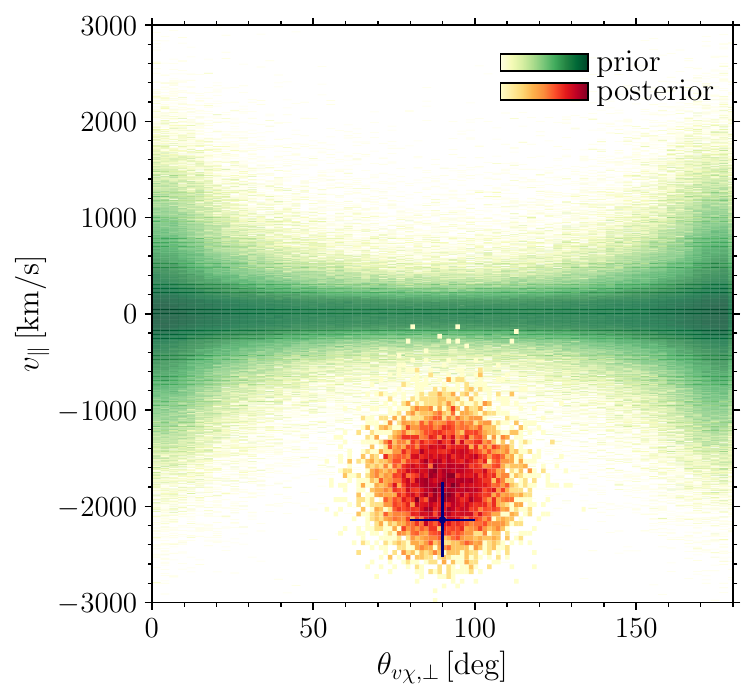}}
        \caption{Prior and posterior distributions for the projection of the kick velocity on the line of sight ($v_{\parallel}$) and the angle between the spin and kick projections onto the plane of the sky ($\theta_{v\chi,\perp}$). The green distribution shows the uninformative prior described in Sect.~\ref{sec:three}. The posterior distribution sampled with our MCMC run is shown in red. Darker (lighter) colors correspond to high (low) density regions. The blue scatter point with error bars indicates the observational constraints where our bi-variate Gaussian likelihood is centered.}
        \label{fig:prior_post}
\end{figure}

\subsection{A rare occurrence?}
\label{subsec:seleff}

As shown in Table \ref{tab:MCMCest}, the inferred value of the recoil velocity along the line of sight $v_{\parallel}\sim 1700$ km/s is systematically lower than the injected median of $2140$ km/s from Eq.~(\ref{musigma}). The low  acceptance rate and large autocorrelation time of the MCMC chains points to an intrinsic difficulty of accommodating the observation of \quasar as a recoiling BH, at least under our uninformative prior assumptions.
As shown in  Fig.~\ref{fig:prior_post}, the likelihood lies in a region of the parameter space where the weight of the prior is vanishingly small --- the two barely overlap.  From our prior distribution, the probability of obtaining $|v_{\parallel} - 2140\, {\rm km/s}|< 390$~km/s is as small as $0.15\%$ while the probability of obtaining $|\theta_{v\chi,\perp} - \ang{90}|< \ang{10}$ is $5.5\%$. When considering both constraints, we find a probability of $1.0\times10^{-3}\%$. The qualitative conclusion one can extract from these numbers is that, if \quasar is a recoiling BH, it must be a rare one.

Selection effects are a crucial caveat to the above statement. Recoiling candidates like \quasar are identified in large spectroscopic surveys because of their highly shifted lines. So the very fact that \quasar was given attention implies that the putative kick is large, and this itself is a rare occurrence. The typical line-shift threshold adopted is of  $\ssim 1000\,\si{km/s}$ \citep{2012ApJS..201...23E}. Selecting only the prior volume with $v_\parallel > 1000\,\si{km/s}$, the fraction of systems compatible with the two observational constraints as above increases to  $\ssim\,0.033\%$.%

More formally, one can quantify the impact of selection biases by computing the Bayes factor between two models in which one does ( or not) account for such selection threshold. In particular, we find
\begin{equation}
        \mathcal{B} = \frac{p(\quasar\,|\,v_{\parallel}\geq1000\,\si{km/s} )}{p(\quasar)}= 31.7\,.
        \label{bayesf}
\end{equation}
This quantifies that, indeed, the probability of detecting \quasar-like systems is enhanced by selection effects in current searches.
We note that the computation of $\mathcal{B}$ from Eq.~(\ref{bayesf}) can be carried out using an appropriate generalization of the so-called Savage-Dickey density ratio and does not require an additional MCMC run.

\begin{figure}[]
        \centering
        \subfloat{\includegraphics[width=0.5\textwidth]{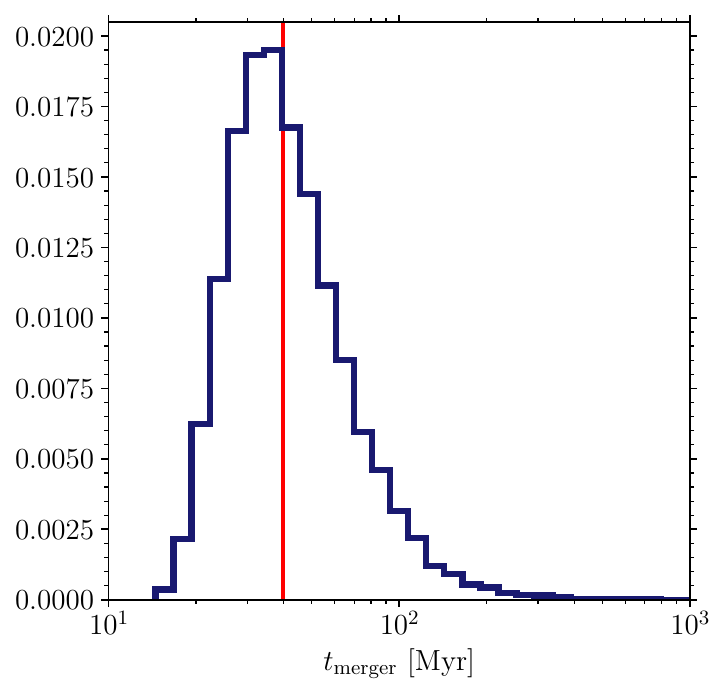}}
        \caption{Elapsed time since the BH merger, as inferred by combining the observed astrometric %
        displacement and the posterior distribution of the projected velocity, $v_{\perp}$. The red line at $\sim 40$ Myr  shows the upper limit of the age of the jet inferred from its synchrotron emission.
        }
        \label{fig:tmerger}
\end{figure}

   \section{Astrophysical implications}
\label{sec:four}

The modeling discussed above sets additional constrains on the nature of \quasar and can be used to identify observational tests of the recoiling hypothesis.

\subsection{Merger time} \label{mergertime}

The reconstruction of the geometry of the putative recoiling BH allows us to place a novel constraint on the elapsed time since the merger of the progenitors. The magnitude of the kick velocity projected onto the plane perpendicular to the line of sight is $v_{\perp}\sim 200\,\si{km/s}$ (see Table \ref{tab:MCMCest}). Astrometric observations measure a displacement of $d_\perp = 11.1\,\si{Kpc} \pm 0.1\,\si{Kpc}$ between the AGN and the isophotal center of the galaxy. We provide a Bayesian measurement of the elapsed time $t_{\rm merger}= d_\perp / v_\perp $ by combining our posterior on $v_{\perp}$  with samples from a normal distribution $p(d_\perp)=  \mathcal{N}(d_\perp\, |\, \mu\!=\!11.1\,\si{Kpc}, \sigma\!=\!0.1\,\si{Kpc})$, thus assuming a constant velocity after the coalescence. Quoting median and 90\% credible interval, we find that the BH merger took place  $50.5^{\,+141.1}_{\,-26.0}\,\si{Myr}$ ago; %
the resulting posterior distribution is shown in Fig.~\ref{fig:tmerger}. %

We next compared the age of the BH remnant to the age of the radio jet. This can be estimated by modeling the spectrum of the emission from the lobes, under the assumption that it is produced by synchrotron emission from electrons accelerated at the shock where the jet collides with the lobe. The age can then be obtained from the position of the ``cooling break'' in the spectrum, which corresponds to the peak of the synchrotron emission from electrons whose cooling time is equal to the age of the source. Assuming the (effectively isotropic, turbulence-driven) magnetic field in the lobe to hold a fraction of order unity of the internal energy in the shocked plasma (``equipartition'')  \citet{1999A&A...345..769M} found $t_\mathrm{age,eq}\sim 0.1\,\mathrm{Myr}$. They note, however, that their equipartition assumption is motivated by  uncertainties on the magnetic field amplification mechanisms, rather than a physical expectation, and that the estimated age of the source scales with the magnetic field strength, $B,$ as
\begin{equation}
 t_\mathrm{age}(B) = t_\mathrm{age,eq}\left(\frac{B}{B_\mathrm{eq}}\right)^{-3/2},
\end{equation}
where $B_\mathrm{eq}=0.7\,\mathrm{mG}$  \citep{1999A&A...345..769M}.
More recent advancements in particle-in-cell simulations of magnetized relativistic collisionless shocks indicate that the downstream magnetic field decays relatively rapidly, reaching energy densities that are orders  of magnitude below those of equipartition in the region where most of the synchrotron emission is produced \citep[e.g.,][]{2015SSRv..191..519S}. This is also supported by observations of synchrotron emission from relativistic shocks caused by gamma-ray-burst-producing relativistic jets \citep{2014ApJ...785...29S}. On the other hand, if the magnetic field were extremely weak, electron cooling in the lobes would have been dominated by inverse Compton scattering with photons of the cosmic microwave background. This effect is equivalent \citep{1999A&A...345..769M} to that of a magnetic field strength $B_{\rm CMB}(z)=3.25(1+z)^2\,\mathrm{\mu G}$, which is equal to $13.4\,\mathrm{\mu G}$ at the redshift $z=1.06$ of \quasar. This poses an upper limit $t_\mathrm{age}\lesssim t_\mathrm{age,eq}(B_\mathrm{CMB}/B_\mathrm{eq})^{-3/2}=3.98\times10^{7}\,\mathrm{yr}\sim 40\,\mathrm{Myr}$. This upper limit accommodates values that are compatible with the merger time $t_{\rm merger}$ derived above (see Fig.~\ref{fig:tmerger}), thus allowing for an interpretation where the jet switched on soon after the BH merger.

If one instead assumes equipartition (i.e., $t_\mathrm{age}\sim 0.1\,\mathrm{Myr)}$, our estimates of $t_{\rm merger}$ implies that \quasar  is a relatively young AGN that switched on tens of megayears after merger. %
In the recoiling scenario, such a delay between the coalescence of the BHs and the triggering of accretion has been predicted by, for example, \cite{2005ApJ...622L..93M}. The time-dependent gravitational potential of the binary carves a gap in the circumbinary accretion disk. Past coalescence, angular-momentum transport processes (modeled with an effective viscosity) refill the gap and restart accretion on the local viscous timescale. \cite{2005ApJ...622L..93M} and \cite{2006MNRAS.372..869D} reported that such a delay in the observer frame should be $\ssim 1.5$ Myr, though it is important to stress that those estimates were calibrated on significantly lighter BHs of $\ssim 10^6$~M$_\odot$. %
While possible, this explanation %
requires some degree of fine-tuning or at least additional modeling to explain what recent event launched the jet.

Selection effects could also be relevant here: if accretion started earlier, it would either be fainter (hence harder to observe, possibly not producing strong broad lines) or exhausted by the time the BH reaches the observed position. Assuming the accretion disk dragged by the recoiling BH is (i) well described by a \cite{1973A&A....24..337S} disk and (ii) limited in its outer radius by self-gravity driven fragmentation, then the total mass available for accretion is $M_{\rm disk} \approx M \times H/R$, where $M$ is the mass of the BH and $H/R\sim 10^{-3}$ is the aspect ratio for a typical AGN accretion disk \citep{2007MNRAS.377L..25K}. A BH accreting close to its Eddington limit would then consume all its fuel on a timescale comparable with the age of the jet, partially motivating its %
youth. %

\subsection{Jet orientation}\label{jetorientation}

\begin{figure}
 \centering
 \includegraphics[width=\columnwidth]{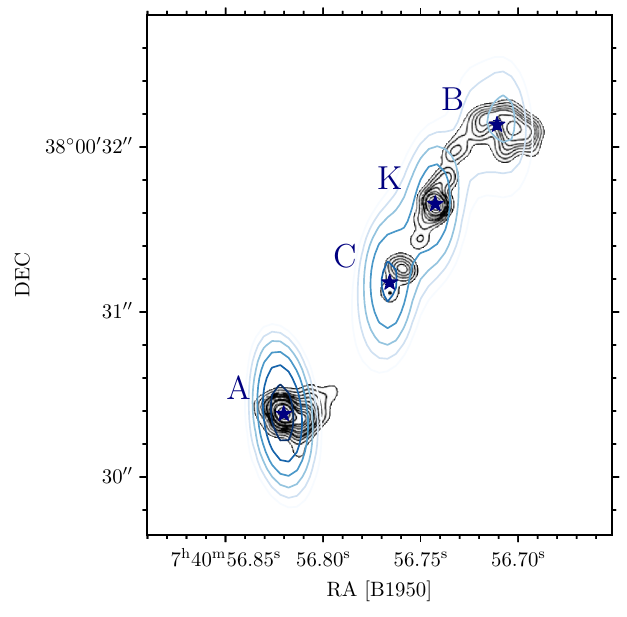}
 \caption{Surface brightness contours of the reconstructed image from archival VLA and EVN+MERLIN observations of \quasar at 8.4 GHz and 1.7 GHz, respectively. Blue curves indicate the VLA observation; we show six contours of constant surface brightness at 8.4 GHz, logarithmically spaced between 10\% and 80\% of the peak surface brightness (33.3 mJy/beam). Black curves in the background refer to the higher resolution EVN+MERLIN image \citep{1991MNRAS.250..225S}. Blue stars mark the centers of four elliptical Gaussian components, which we label A (lobe of the receding jet), B (lobe of the approaching jet), C (core), and K (knot).}
 \label{fig:VLA+EVN_map}
\end{figure}

We assumed that the jet axis is aligned with the BH spin, so the jet viewing angle (i.e., the angle between the line of sight and the jet axis) is
\begin{equation}
 \theta_\mathrm{jet}=\left\lbrace\begin{array}{ll}
                                \theta_{n\chi} & \;\;{\rm if}\qquad 0^\circ\leq \theta_{n\chi}<90^\circ\,,\\
                                180^\circ-\theta_{n\chi} & \;\;{\rm if}\qquad 90^\circ\leq \theta_{n\chi}<180^\circ\,.\\
                               \end{array}\right.
\end{equation}
From our MCMC we find $\theta_\mathrm{jet} = {7.2^\circ}_{-\,5.2^{\circ}}^{+\,9.8^{\circ}}$. %
The jet viewing angle can also be measured using radio data and, in the following, we present two different strategies to do so.

We reduced and analyzed the VLA data of \quasar using version 6.5.4 of the \textsc{casa} software \citep{2022PASP..134k4501C}. In particular, we calibrated the flux density scale using the close-by 0812+367 calibrator as reference and we reconstructed the image %
using natural weighting and a 50 mas cell size. Constant surface brightness contours of the resulting image are shown in blue in Fig.~\ref{fig:VLA+EVN_map}. Blue stars mark the centers of four elliptical Gaussian components that we fitted to the image, %
indicated by A, B, C, and K. %
In Fig.~\ref{fig:VLA+EVN_map} we superimpose in black the isocontours of surface brightness obtained from the more sensitive, higher-resolution observation performed with EVN+MERLIN at 1.7 GHz \citep{1991MNRAS.250..225S}.  Coordinates in the figure refer to the VLA observation and the EVN+MERLIN contours have been aligned using the position of component A. %

Consistently with \citet{1991MNRAS.250..225S} and given the spectral information reported there, we identify the following features.

\begin{itemize}
\item We interpret component C as the jet ``core,'' whose projected position should be consistent with that of the supermassive BH to within a few milliarcseconds \citep{2011A&A...532A..38S}.
\item The component labeled with K is the ``knot'' of the jet, marking dissipation of energy by means of, for example,\ internal shocks \citep{1978MNRAS.184P..61R}.
\item Components A and B are the ``lobes,'' where the  kinetic energy of two oppositely oriented relativistic jets is converted into internal energy due to a strong shock caused by collision of the jet particle with the interstellar medium \citep{1974MNRAS.169..395B}. 
\end{itemize}
Given the absence of visible emission between the core and lobe A, we interpret lobe A to be associated with the receding jet (or ``counterjet'') and lobe B to be associated with the approaching jet.

\subsubsection{Approaching and receding jet brightness ratio}
\label{sec:brightness_ratio}

The first method for estimating  $\theta_\mathrm{jet}$ is based on the observed brightness ratio  $\mathcal R $ between the approaching and receding jets. Assuming Doppler boosting of otherwise identical jets, one has \citep{1979ApJ...232...34B,1979Natur.277..182S}
\begin{equation}\label{eq:DB}
 \mathcal R =\frac{F_\mathrm{\nu,a}}{F_\mathrm{\nu,r}}=\left(\frac{1+\beta\cos \theta_\mathrm{jet}}{1-\beta\cos \theta_\mathrm{jet}}\right)^{2+\alpha},
\end{equation}
where $F_\mathrm{\nu,a}$ indicates the flux density of the approaching jet, which we identify with that of component K, and $F_\mathrm{\nu,r}$ is the flux density of the receding jet, which is undetected. The spectral index $\alpha$ is defined such that $F_\mathrm{\nu} \propto \nu^{- \alpha}$ for both jets; for component K \citet{1991MNRAS.250..225S} reports $\alpha=1.2$. The jet bulk velocity in units of the speed of light is given by $\beta=(1-\gamma^{-2})^{1/2}$, where $\gamma$ is the associated Lorentz factor.

The posterior probability on the viewing angle and the Lorentz factor given the surface brightness data, $\vec d,$ from the image is given by
\begin{equation}
 p(\theta_\mathrm{jet},\gamma\,|\,\vec d) = \frac{p(\vec d\,|\,\theta_\mathrm{jet},\gamma)\pi(\theta_\mathrm{jet})\pi(\gamma)}{p(\vec d)},
 \label{eq:post1}
\end{equation}
where $\pi(\theta_\mathrm{jet})$ and $\pi(\gamma)$ are prior probabilities and the likelihood can be written as \begin{align}
 p(\vec d\,|\,\theta_\mathrm{jet},\gamma)= \iint p(\vec d\,|\,F_\mathrm{\nu,a},F_\mathrm{\nu,r})p(F_\mathrm{\nu,a},F_\mathrm{\nu,r}\,|\,\theta_\mathrm{jet},\gamma)\,&\mathrm{d}F_\mathrm{\nu,r}\mathrm{d}F_\mathrm{\nu,a} 
 \notag \\
= \iint p(\vec d\,|\,F_\mathrm{\nu,a},F_\mathrm{\nu,r})
 \delta[F_\mathrm{\nu,a}-\mathcal{R}(\theta_\mathrm{jet},\gamma)F_\mathrm{\nu,r}]\pi(F_\mathrm{\nu,r})
 \,&\mathrm{d}F_\mathrm{\nu,r}\mathrm{d}F_\mathrm{\nu,a}\,,
 \label{eq:like1}
\end{align}
where %
where the second equality follows from $\theta_\mathrm{jet}$ and $\gamma$ being uninformative with respect to either of the flux densities alone, while informing their ratio through Eq.~(\ref{eq:DB}).
The likelihood $p(\vec d\,|\,F_\mathrm{\nu,a},F_\mathrm{\nu,r})$ can be expressed %
as
\begin{equation}
 p(\vec d\,|\,F_\mathrm{\nu,a},F_\mathrm{\nu,r}) = \frac{p(\vec F_\mathrm{\nu,r}\,|\,\vec d)p(\vec F_\mathrm{\nu,a}\,|\,\vec d)}{\pi(F_\mathrm{\nu,a})\pi(F_\mathrm{\nu,r})}p(\vec d).
 \label{eq:like2}
\end{equation}
Combining Eqs.~(\ref{eq:post1}), (\ref{eq:like1}), and (\ref{eq:like2}) and carrying out the integral on $F_\mathrm{\nu,a}$ yields 
\begin{equation}
 p(\theta_\mathrm{jet},\gamma\,|\,\vec d) = \pi(\theta_\mathrm{jet})\pi(\gamma)\int \mathcal{L}_{F_\mathrm{\nu,a}}\left[\mathcal{R}(\theta_\mathrm{jet},\gamma)F_\mathrm{\nu,r}\right]p(F_\mathrm{\nu,r}\,|\,\vec d)\,\mathrm{d}F_\mathrm{\nu,r},
 \label{informed}
\end{equation}
where we have defined
\begin{equation}
 \mathcal{L}_{F_\mathrm{\nu,a}}(x)=\left.\frac{p(F_\mathrm{\nu,a}\,|\,\vec d)}{\pi(F_\mathrm{\nu,a})}\right|_{F_\mathrm{\nu,a}=x}.
\end{equation}
In words, Eq.~(\ref{informed}) is the joint posterior probability of the viewing angle and the bulk Lorentz factor informed by the approaching and receding jet flux density measurements. We can then marginalize this expression over $\gamma$ to obtain the posterior probability $p(\theta_{\rm jet}| \vec d )$. %
In practice, we approximated these integrals using Monte Carlo sums.
 
Our %
Gaussian %
fit of the VLA  radio %
data
 yields a total flux density for the K component of $F_\mathrm{\nu,K}=21.4\pm1.3\,\mathrm{mJy}$, where we report one-sigma confidence intervals. Based on this, we approximate %
\begin{equation}
 \mathcal{L}_{F_\mathrm{\nu,a}}(x)\propto \mathcal{N}(x \,|\, \mu\!=\!21.4\,\mathrm{mJy},\sigma\!=\!1.3\,\mathrm{mJy}).
 \end{equation}
At the putative position of the receding jet, that is, at the mirror position of knot K with respect to the core C,
the surface brightness in our VLA image is $I_\mathrm{\nu,r}\sim 0.5\,\mathrm{mJy/beam}$. Assuming the image noise to be reasonably well modeled as a zero-mean Gaussian with standard deviation equal to the measured root-mean-square noise $\sigma_\mathrm{rms}=0.6\,\mathrm{mJy/beam}$ in the image, assuming the receding jet to be un-resolved, and adopting a uniform prior on its flux density, one gets
\begin{equation}
 p(F_\mathrm{\nu,r}\,|\,\vec d)\sim \left(2\pi\sigma_\mathrm{rms}^2\right)^{-1/2}\exp\left[-\frac{1}{2}\left(\frac{F_\mathrm{\nu,r}/\theta_\mathrm{b}^2-I_\mathrm{\nu,r}}{\sigma_\mathrm{rms}}\right)^2\right],
\end{equation}
where $\theta_\mathrm{b}^2$ is the beam size, which does not need be specified as it simplifies. %
We adopted a mildly informative, scale-free prior probability  on the bulk jet Lorentz factor   $\pi(\gamma)$ that is uniform in log between 2 and 30. %
This range accommodates values of the Lorentz factor estimated from superluminal motion in AGN jets \citep{1994ApJ...430..467V}.

\begin{figure}
 \centering
 \includegraphics[width=0.5\textwidth]{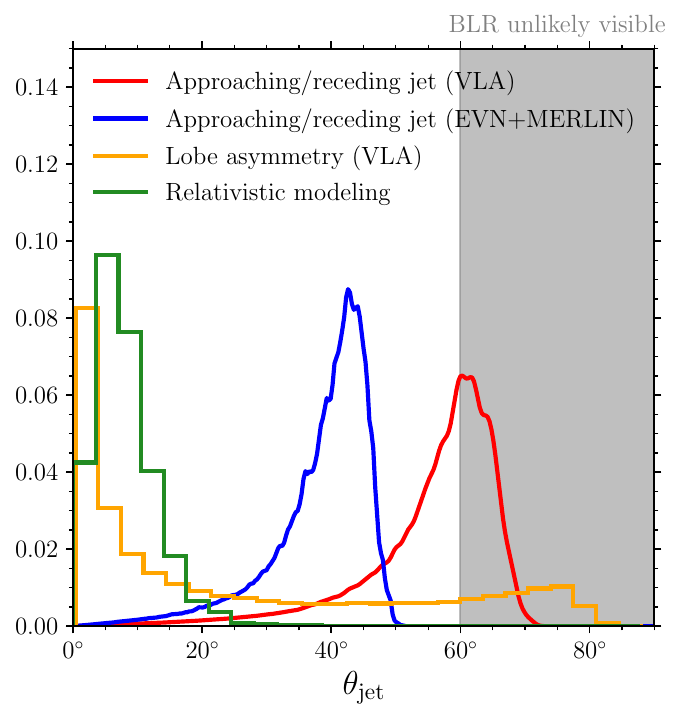}
 \caption{Posterior distribution of the jet viewing angle, $\theta_{\rm jet}$. The red (blue) curve shows results from the approaching and receding jet brightness method (Sect.\ \ref{sec:brightness_ratio}) using data from archival VLA (EVN+MERLIN) observations. The yellow histogram shows results from the lobe asymmetry method (Sect.\ \ref{sec:lobe_asymmetry}), adopting a log-uniform prior on the source age in the range $0.1\leq t_\mathrm{age}/\mathrm{Myr}\leq 40$. The green histogram shows the posterior on $\theta_{\rm jet}$ from our relativistic model (Sect. \ref{sec:three}). The gray shaded region corresponds to viewing angles that are disfavored by the observations of broad lines (BLR) in the optical spectrum of the core. %
 }
 \label{fig:thv_posteriors}
\end{figure}

The red distribution in Fig.~\ref{fig:thv_posteriors} shows the resulting posterior on the jet viewing angle $\theta_\mathrm{jet}$. Since the measurement of the receding jet flux density is not constraining, the result favors relatively large viewing angles owing to the fact that these are sufficient to produce mild flux-density contrasts. %
We find that  $\sim 42\%$ %
of the posterior weight lies in the region where $\theta_\mathrm{jet}>60^\circ$, which is actually disfavored since the broad line region is unobscured \citep[e.g.,][]{1995ApJ...452L..95B}.
For comparison, the blue distribution shows an analogous result obtained the higher-resolution observations by \cite{1991MNRAS.250..225S}. In particular, we estimated $F_\mathrm{\nu,K}=80\pm 0.15\,\mathrm{mJy}$, where the central value is reported in the cited paper and  the error is set equal to the image noise rms, and $F_\mathrm{\nu,K}=0.15\pm 0.15\,\mathrm{mJy}$ for the receding jet, as the corresponding feature is not visible in the image. While  higher-sensitivity data result in a tighter constraint on the viewing angle, neither of the two datasets allows for a precise determination.

\subsubsection{Lobe asymmetry}
\label{sec:lobe_asymmetry}

Our second method for estimating $\theta_{\rm jet}$ relies on the fact that the receding lobe (component A) is located at a slightly larger distance from Earth than the approaching lobe (component B). The gap between the two must have been increasing over time as the lobes were pushed farther out by the jets. Photons reaching us from A have thus been traveling for longer than those from B, and this produces an asymmetry in the apparent lobe separations from the core, which is in principle measurable. We assumed the two lobes moved with the same average speed, $\beta_\mathrm{L}$ (in units of $c$), from their initial position, which was presumably close to the core. We also assumed that the two jets have similar properties and that the background inter-galactic medium is the same on both sides of the core.

The ratio of the lobe separations $s_\mathrm{A}$ and $s_\mathrm{B}$ from the core is given by \citep{1995MNRAS.277..331S}

\begin{equation}
 \frac{s_\mathrm{B}}{s_\mathrm{A}}=\frac{1+\beta_\mathrm{L}\cos\theta_\mathrm{jet}}{1-\beta_\mathrm{L}\cos\theta_\mathrm{jet}},
\end{equation}
so
\begin{equation}
 \beta_\mathrm{L}\cos\theta_\mathrm{jet} = \frac{s_\mathrm{B}/s_\mathrm{A}-1}{s_\mathrm{B}/s_\mathrm{A}+1}.
\label{eq:proj_lobe_speed1}
 \end{equation}
The apparent speed of the lobes is
\begin{equation}
 \beta_\mathrm{app,A} = \frac{\beta_\mathrm{L}\sin\theta_\mathrm{jet}}{1+\beta_\mathrm{L}\cos\theta_\mathrm{jet}}, \qquad \beta_\mathrm{app,B} = \frac{\beta_\mathrm{L}\sin\theta_\mathrm{jet}}{1-\beta_\mathrm{L}\cos\theta_\mathrm{jet}},
\end{equation}
such that 
\begin{equation}
 \beta_\mathrm{app,rel}=2\tan\theta_\mathrm{jet}\frac{\beta_\mathrm{L}\cos\theta_\mathrm{jet}}{1-(\beta_\mathrm{L}\cos\theta_\mathrm{jet})^2}
 \end{equation}
is their apparent relative speed.
After a time $t_\mathrm{age}$ (as measured in the rest frame of the core), the apparent projected physical separation between the lobes is  $s_\mathrm{AB} d_\mathrm{A} = \beta_\mathrm{app,rel} ct_\mathrm{age}$
where $s_\mathrm{AB}$ is the separation between the lobes %
and $d_\mathrm{A} = 35.38$ kpc/arcsec %
is the angular diameter distance at the redshift of \quasar, assuming a flat Friedmann-Lema\^itre-Robertson-Walker universe with cosmological parameters from \citet{2016A&A...594A..13P}.
Putting the pieces together, we find
\begin{equation}
 \beta_\mathrm{L}\cos\theta_\mathrm{jet}=\frac{c t_\mathrm{age} \tan\theta_\mathrm{jet}}{s_\mathrm{AB}d_\mathrm{A}}\left[\sqrt{1+\left(\frac{s_\mathrm{AB}d_\mathrm{A}}{c t_\mathrm{age} \tan\theta_\mathrm{jet}}\right)^2}-1\right]\,.
\label{eq:proj_lobe_speed2}
\end{equation}

Combining Eqs.~(\ref{eq:proj_lobe_speed1}) and (\ref{eq:proj_lobe_speed2}) results in an implicit equation that relates the viewing angle $\theta_\mathrm{jet}$ to quantities that can be measured from the image, namely $s_\mathrm{A}$, $s_\mathrm{B}$, and $s_\mathrm{AB}$, as well as the source age $t_\mathrm{age}$. For small degrees of asymmetry $s_{\rm B}/ s_{\rm A}\to 1$, %
the expression can be simplified and solved analytically, leading to
\begin{equation}
 \tan\theta_\mathrm{jet}\sim\left(\frac{s_\mathrm{AB}d_\mathrm{A}}{2ct_\mathrm{age}}\right)\frac{s_\mathrm{B}/s_\mathrm{A}+1}{s_\mathrm{B}/s_\mathrm{A}-1},
\end{equation}
but the solution to the full equation can be easily obtained numerically. The results reported below make use of the numerical solution.%

Using VLA data, we measure $s_\mathrm{A}=1.023\pm 0.011\,\mathrm{arcsec}$, $s_\mathrm{B}=1.157\pm 0.025\,\mathrm{arcsec}$, and $s_{AB}=2.178\pm 0.023\,\mathrm{arcsec}$, where we quote means and 1$\sigma$ intervals. %
We assumed a prior on $t_\mathrm{age}$ that is uniform in log between 0.1 Myr and 40 Myr, which accommodates the uncertainties detailed in Sect.~\ref{mergertime}.  We thus estimated the jet viewing angle $\theta_{\rm jet}$ by sampling $s_\mathrm{A}$, $s_\mathrm{B}$, and $s_\mathrm{AB}$ %
 from Gaussian distributions (with means and standard deviations corresponding to the estimates reported above), sampling  $t_\mathrm{age}$ from its prior, and evaluating the numerical solution for $\theta_{\rm jet}$ described above.
The resulting distribution is shown in orange in Fig.~\ref{fig:thv_posteriors}. Because of the small degree of lobe asymmetry observed and  the ``old ages'' of $\mathcal{O}(10)$ Myr  allowed by the prior (hence relatively low lobe speeds), this estimate  favors small viewing angles %
($\theta_{v}<69.9^{\circ}$ at 90\% confidence), which are in good agreement with the recoiling scenario.

      \section{Conclusions}
\label{sec:five}

Black hole merger kicks are a direct consequence of linear-momentum dissipation in strong-field gravity. \quasar~--- a quasar exhibiting both spatial and velocity displacements from its host galaxy --- is arguably the most promising astrophysical candidate to date \citep{2017A&A...600A..57C}. The source also shows a prominent radio jet with several resolved features \citep{1991MNRAS.250..225S}.

When modeled as a recoiling BH in general relativity, we find that there exists a region in the parameter space of the possible progenitor binaries that explains all the pieces of observational evidence at our disposal, namely the measured line shift \citep{2017A&A...600A..57C} and the projected direction of the jet \citep{2022ApJ...931..165M}.
The previous analysis by \cite{2017ApJ...841L..28L} considered progenitor binaries resulting in a remnant recoiling with an (intrinsic, not projected) velocity $v>2000$ km s$^{-1}$. Using their model with a uniform prior and quoting 1$\sigma$ errors, they report a preference for progenitor binaries with mass ratios $q = 0.49^{+0.26}_{-0.18}$ and spins that are close to unity $\bigl(\chi_1 = 0.99^{+0}_{-0.46}$, $\chi_2 = 0.99^{+0}_{-0.26}\bigr)$ and misaligned with respect to the orbital angular momentum. As a consequence of the large recoil, they also infer that the imparted kick is nearly collinear with the orbital angular momentum, which is expected as their study predates the consideration on the jet direction by \cite{2022ApJ...931..165M}. When taking projection effects into account, we find $v=1759^{+581}_{-643}$ km s$^{-1}$, which indicates that the threshold imposed by \cite{2017ApJ...841L..28L} might be too restrictive. For a comparison with the above values, our best estimates are $q=0.57^{+0.37}_{-0.29}$, $\chi_1=0.89^{+0.10}_{-0.41}$, and $\chi_2=0.60^{+0.37}_{-0.54}$ at 90\% credibility.

In particular, we fully reconstruct the geometry of the system, showing that both the spin and the kick vectors must be pointing toward or away from us. However, the recoiling BH scenario requires some amount of fine-tuning, at least under uninformative prior assumptions.
We conclude that, if \quasar is indeed a recoiling BH, it must be a rare one. 

Such a low probability of obtaining a configuration similar to that of \quasar %
implies the existence of a much larger population of recoiling BHs with projected velocities $v_{\parallel} \gtrsim 1000$ km s$^{-1}$ and redshifts  $z\approx 1$ within the Sloan Digital Sky Survey footprints. They should be the remnants of close-to-equal-mass binaries of $\gtrsim
\! 10^9$ M$_\odot$ that merged within the last $\ssim 50$ Myr. 
If all these systems were to be accreting, the expected number would be far larger than the tens of systems identified within the Sloan Digital Sky Survey presenting large velocity shifts between the broad and narrow emission lines \citep{2011ApJ...738...20T, 2012ApJS..201...23E}. An important caveat to this statement is that %
the majority of such high-velocity recoiling BHs could be inactive. %
For future work, our modeling can be used to compute the expected merger rate of such \quasar-like objects and check whether it is compatible with current GW constraints from pulsar timing arrays \citep{2023arXiv230616227A,2023ApJ...952L..37A} as well as the rate of massive galaxy mergers~\citep{2015A&A...576A..53L}.

We then attempted an observational verification of our findings using archival radio data. First, our relativistic model indicates that the BH merger must have happened $\mathcal{O}(10)$ Myr ago. We believe this is compatible with the age inferred from the emission spectrum of the radio lobe, provided one appropriately rescales the nominal age  \citep{1999A&A...345..769M} with the expected magnetic field. Second, assuming the radio jet is launched along the direction of the BH spin, our modeling predicts that the viewing angle of the jet must be small.  
We tested this using both the spatial and the flux asymmetry in the approaching and receding jets, and found that our constraining power is ultimately limited by the quality of the available archival data. %
Deeper radio observations of \quasar might shed light on its nature as a recoiling supermassive~BH.

\begin{acknowledgements}
We thank Vishal Baibhav, Emanuele Berti,
Marco Chiaberge, David Izquierdo-Villalba, Colin Norman, Alberto Sesana, Ralph Spencer, David Williams and Georg Herzog for discussions. M.B. and D.G. are supported by ERC Starting Grant No.~945155--GWmining,
Cariplo Foundation Grant No.~2021-0555, MUR PRIN Grant No.~2022-Z9X4XS, 
and the ICSC National Research Centre funded by NextGenerationEU. 
D.G. is supported by MSCA Fellowships No.~101064542--StochRewind and No.~101149270--ProtoBH. 
O.S.S.\ acknowledges support by the PRIN-INAF grant 1.05.23.04.04.
Computational work was performed at CINECA with allocations 
through INFN and Bicocca.

\end{acknowledgements}

\bibliographystyle{aa_edited}
\bibliography{3c186}

\clearpage
\onecolumn
\begin{appendix}

\section{Additional runs}
\label{secapp}

\begin{figure*}[hb]
        \centering
        \subfloat{\includegraphics[width=\textwidth]{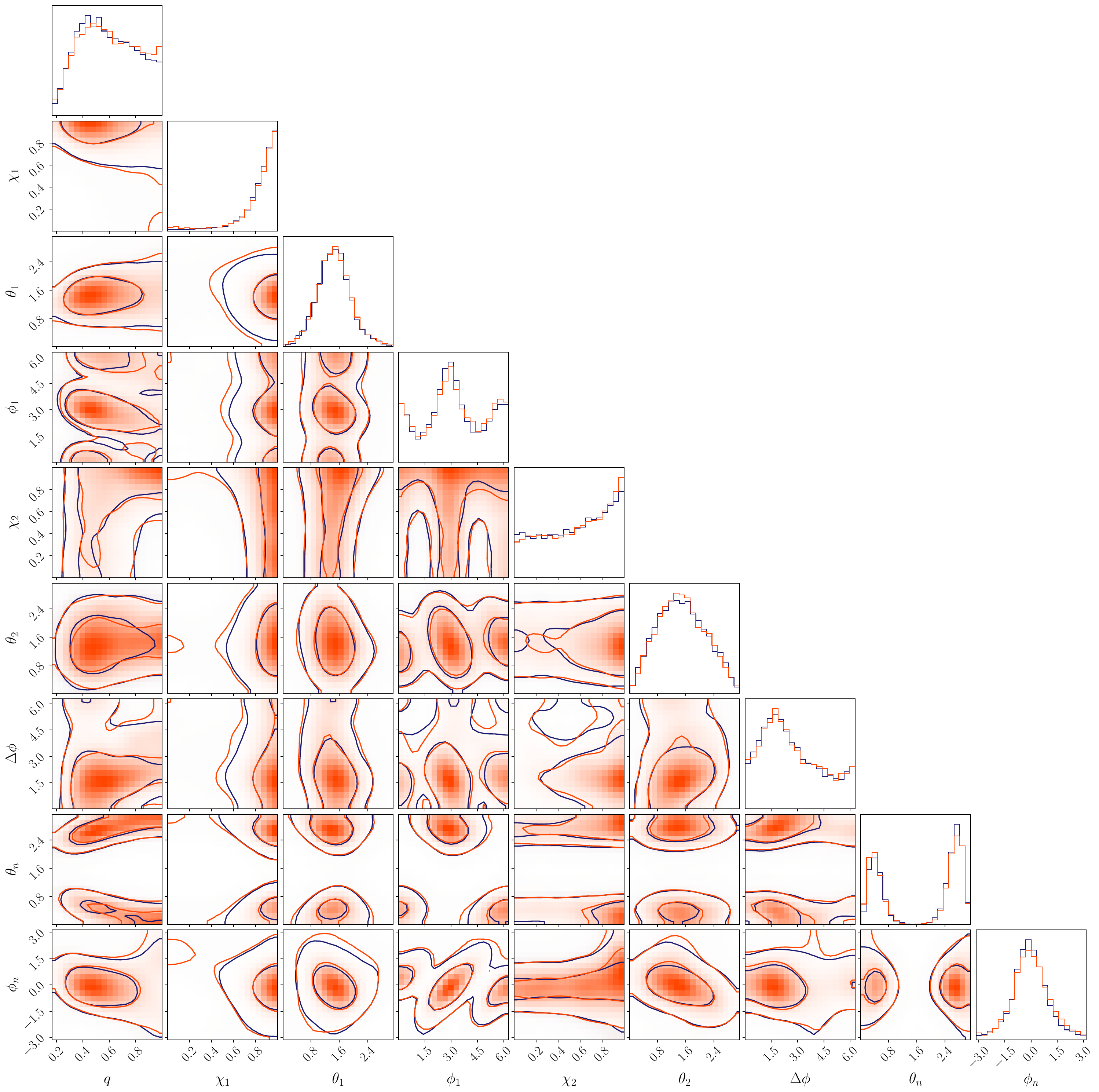}}
        \caption{Joint posterior distributions of the parameters describing the binary progenitors of \quasar assuming a $\ang{10}$ (blue) or $\ang{1}$ (red) error on $\theta_{v\chi,\perp}$. The blue distribution is the same as in Fig.~\ref{fig:cornerplot}. 
        } 
        \label{fig:cornerplot3}
\end{figure*}

\begin{figure*}[t]
        \centering
        \subfloat{\includegraphics[width=\textwidth]{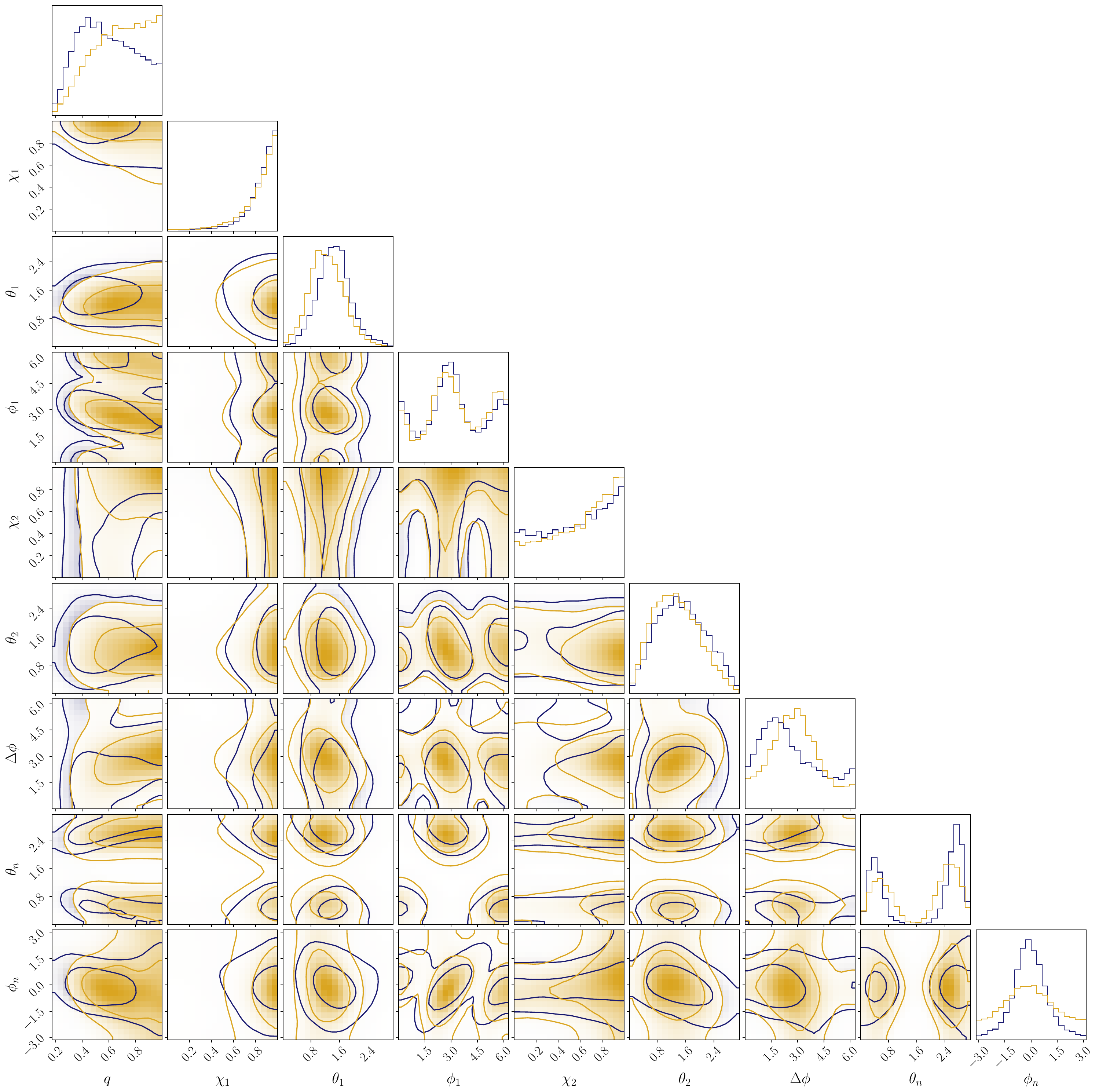}}
        \caption{Joint posterior distributions of the parameters describing the binary progenitors of \quasar with (blue) and without (gold) the constraint on $\theta_{v\chi,\perp}$. The blue distribution is the same as in Fig.~\ref{fig:cornerplot}. 
        } 
        \label{fig:cornerplot2}
\end{figure*}

We present results from some additional MCMC runs, extending what is presented in Sect.~\ref{sec:three}. 
In Fig.~\ref{fig:cornerplot3} we repeat our analysis assuming the standard deviation of $\theta_{v\chi,\perp}$ is $\ang{1}$ instead of the more conservative value $\ang{10}$ used in the many body of the paper. The results are indistinguishable.
In Fig. \ref{fig:cornerplot2} we ignore the constraint on the projected spin-kick angle ($\theta_{v\chi,\perp} = \ang{90} \pm \ang{10}$) and only consider that on the projected kick velocity ($v_{\parallel} = -2140 \pm 390\,\si{km/s}$).
In this case the posterior is different, with the most affected parameters being the mass ratio ($q$), the angle between the in-plane spin projection ($\Delta\phi$), and the orientation angles ($\theta_n,\phi_n$).

\end{appendix}
\end{document}